\documentclass[aps,prl,twocolumn,showpacs,superscriptaddress,groupedaddress,floatfix]{revtex4-1}
\usepackage{graphicx}
\usepackage{amsmath}
\usepackage{amssymb}
\usepackage{dcolumn}
\usepackage{natbib}
\usepackage{dsfont}
\usepackage{latexsym}
\usepackage{rotating}
\usepackage{color}
\usepackage{latexsym}
\usepackage{bbm}
\usepackage{subfigure}
\usepackage{float}
\usepackage{epsfig}
\usepackage{psfrag}
\usepackage{bm}
\usepackage{amsthm}
\usepackage{eucal}
\usepackage{mathrsfs}
\usepackage{braket}
\usepackage{psfrag}
\usepackage{enumerate}
\usepackage{longtable}
\usepackage{subfigure}
\usepackage{bm}
\usepackage{hyperref}
\hypersetup{
colorlinks=true,final=true,
        linkcolor=red,
        citecolor=blue,
        filecolor=blue,
        urlcolor=blue,
}

\usepackage{dcolumn}
\usepackage{bm}
\usepackage{dcolumn}
\usepackage{bm}
\newcommand{\be}{\begin{equation}}
\newcommand{\ee}{\end{equation}}
\newcommand{\bn}{\begin{eqnarray}}
\newcommand{\en}{\end{eqnarray}}
\begin{document}

\title{Emergent Quantum Coherence from Instabilities of a Perturbed Fractionalized Spin Liquid}

\author{S. Acharya$^{1}$}\email{swagata@phy.iitkgp.ernet.in}
\author{M. S. Laad$^{2}$}\email{mslaad@imsc.res.in} 
\author{A. Taraphder$^{1,3}$}\email{arghya@phy.iitkgp.ernet.in}
\title{Emergent Quantum Coherence from Instabilities of a Perturbed Fractionalized Spin Liquid}
\affiliation{$^{1}$Department of Physics, Indian Institute of Technology, 
Kharagpur, Kharagpur 721302, India.}
\affiliation{$^{2}$Institute of Mathematical Sciences, Taramani, Chennai 600113, India
}
\affiliation{$^{3}$Centre for Theoretical Studies, Indian Institute of 
Technology Kharagpur, Kharagpur 721302, India.}


\begin{abstract}
{
  Effects of perturbations
on topologically ordered (TO) states with fractionalized excitations is very intriguing.  
While small perturbations are not expected to affect a robust TO, sizable 
perturbations must lead to onset of novel routes to order(s).  Observability of such emergent orders 
in real condensed matter systems continues to be an open issue of paramount interest.
Here, we detail how a featureless Kitaev spin liquid with rigorous TO
gives way to partial topological ordered state(s) coexisting with a 
wide staircase of novel `excitonic' orders of fractionalized entities in a Zeeman field. 
We provide quasi-exact and novel
realization for a range of strange features, from nodal spin-metals, 
fractionalization in selective-Mott states, to Luttinger surfaces and 
hidden coherence, all characteristics of the influential resonating-
valence-bond as well as fractionalized-Fermi-Liquid  
views in the context of strange metals. As a novel application, we show 
how Josephson-junction array may exhibit fractional Josephson oscillations upon flux 
tuning as a particularly novel manifestation of emergent quantum coherence.
}
\end{abstract}

\maketitle

  Topological excitations have fascinated physicists for many decades.  
Remarkably, recent advances in condensed matter have reached a stage where the 
ability to observe and even manipulate such exotic excitations seem to be 
within the realm of possibility. Cold-atom set-ups, topological insulators and 
appropriately engineered Josephson junction arrays (JJA) are poised to offer renewed insight into topological excitations arising from ground states with topological order (TO).
In many instances, however, ensuring their stability is a potential hurdle.  
Another, more exacting hurdle is to 
unearth specific examples of systems capable of exhibiting stable (symmetry 
protected) TO.

The birth of the exactly soluble Kitaev model (KM)~\cite{kitaev} led to a spurt of proposals to `simulate' 
its TO ground state in various contexts~\cite{jila,jja}.  In contrast, a microscopic understanding of effects of various perturbations is lacking: while mostly studied numerically~\cite{trebst}, 
a coherent picture of their effects in terms of changes in the spectrum of topological (elementary) 
excitations of the TO state has remained elusive.  Given that the KM hosts a  
fractionalized spin liquid {\it phase}, novel, unanticipated ordered 
states (partial or complete deconfinement-confinement transition(s)), depending upon specific nature of perturbations, are expected 
theoretically. Classifying such novel orders aids the quest for their 
observability in natural or engineered condensed matter systems, apart from the 
attractive motivation of application(s) to fault-tolerant quantum computing. 
Here, we are motivated by recent proposals in 
perturbed Kitaev models constructed by suitably engineered JJA~\cite{jja}.  Moreover, for the realistic range of 
physical parameters, such JJA (see Fig.(1) of supplementary informations) are {\it always} 
perturbed by `magnetic fields' along $x$ and/or $z$ bonds.

 Hence, they offer an attractive avenue to investigate the above issues in 
a realistic context.  Further, models that are variations of the KM and compass models~\cite{JvdB} are
believed to apply even to certain transition-metal oxides (TMO)~\cite{giniyat}, though, in this context, 
additional perturbation(s) must be included~\cite{ybkim,imada}. 
In this context, fascinating issues relate to whether a fractionalized orbital liquid can 
{\it emerge}, and whether analogues of {\it orbital}-Kondo destroying QCPs exist in TMOs.   
\vspace{0.2cm}

\noindent {\bf A Simple Perturbed Kitaev Model}
\vspace{0.2cm}

Consider the Kitaev model in an external Zeeman magnetic field:

\be
H=-\sum_{\alpha}J_{\alpha}\sum_{<i,j>}S_{i}^{\alpha}S_{j}^{\alpha}-
h_{z}\sum_{i}S_{i}^{z}
\ee

\noindent Using the $D=2$ Jordan-Wigner mapping~\cite{hu}, $H$(see supplementary informations\footnote{For supplementary informations please check the arXiv link.}) can be brought to the following
instructive form (this also bares the {\it exact} emergent low-dimensional 
gauge symmetries, as pointed out by Fradkin et al.~\cite{fradkin}):
$H=H_{K}+H_{z}$, where 

\begin{align}
H_{K}=&\sum_{q}[\epsilon_{q}c_{q}^{\dag}c_{q}+\frac{i\Delta_{q}}{2}(c_{q}^{\dag}c_{-q}^{\dag}+h.c)]
\nonumber\\&+\frac{J_{z}}{4}\sum_{i}(2n_{i,\alpha}-1)(2c_{i}^{\dag}c_{i}-1)
\end{align}
and $H_{z}=2h_{z}\sum_{i}(c_{i}^{\dag}\alpha_{i}+h.c)$.
This is just a spinless Falicov-Kimball model with a local 
hybridization $h_{z}$ in $d=2$, but with an additional `spin'-triplet pairing 
term, whence topologically ordered ground state of $H_K$ now appears as 
a spin-triplet BCS superfluid of JW fermions. Here, 
we define $\alpha_{i}'=(2n_{i\alpha}-1)=iB_{i1}B_{i2}=
\pm 1$, which is now a {\it dynamic} gauge field variable on the center of 
each zz-links of the honeycomb Kitaev model (when $n_{i,\alpha}=0,1$ for all $i$. 
This transforms the original 
spin model on a honeycomb lattice to a two- ``orbital" spinless model on an 
effective square lattice. While $[n_{i\alpha},H]=0$ for all $i$ (reminiscent of 
the spinless FK model) ensures the {\it exact} solvability of the KM when 
$h_{z}=0$, a finite Zeeman field immediately spoils the beautiful integrability 
of $H$. Interestingly, however, finite $h_{z}$ simply translates to a local 
hybridization between the $c, \alpha$ JW fermions. These observations are 
crucial for an `almost exact' analysis of the KM in a Zeeman field 
as detailed below.

  Alternatively, we motivate the 
instabilities of the TO state of $H_{K}$ as follows. In Kitaev's language, a 
partial local $Z_{2}$ symmetry associated with $[b_{i}^{a}b_{j}^{a}, 
H_{K}]=0$ for all $(ij)$ and $a=x,y$ still holds when $h_{z}\neq 0$ 
(only $[b_{i}^{z}b_{j}^{z},H_{K}]\neq 0$ along $zz$-links, since 
$H_{z}=i\sum_{i}b_{i}^{z}c_{i}$ mixes $b^{z},c$ on each site). This directly 
implies a {\it dimensional reduction} and associated lower-$D(=d<D)$ (here, $d=1$) 
gauge-like symmetries (GLS) naturally emerge. The only {\it order} that survives 
the action of these GLS is a directional spin-{\it nematic} order~\cite{fradkin}, 
whose order parameter, $\langle n\rangle=\langle S_{i}^{x}S_{i+e_{x}}^{x}-S_{i}^{y}S_{i+e_{y}}^{y}\rangle$ is the only one
that remains invariant under the emergent $d=1$ GLS: thus, the TO of $H_{K}$ is only partially ``melted'' by a Zeeman field.  The ultra-short-ranged spin-liquid phase
known for $h_{z}=0$ thus immediately becomes unstable to a {\it deconfined} critical liquid inherently unstable to incipient spin-nematic order (on xx,yy zig-zag links) co-existing with a  (with finite flux on zz-links) field-induced zz-spin order, for any $h_{z}\neq 0$ at $T=0$!  It also turns out that the $zz$-spin correlation function, $\chi_{zz}(i-j)\simeq (i-j)^{-4}$ (not shown, proof essentially identical to Tikhonov {\it et al.}~\cite{feigelman}) using a simple bubble composed of the $G_{b^{z}b^{z}}(i-j), G_{cc}(i-j)$ propagators for $H$ in Majorana language.  Remarkably, field-induced magnetization now appears as an excitonic condensate of JW-fermions. 

  In JW-fermion language, for special values of $h_{z}$, the now {\it dynamical} $Z_{2}$ 
flux can, by itself, self-organize into stripe-ordered patterns, in analogy with crystal states found in the 
FK model~\cite{lieb} (here, $n_{i\alpha}=(1+\alpha_{i})/2$ play the role of immobile fermions in 
the FK-like model for $h_{z}=0$). In the present case, such crystals occurring with a large range of 
periodic structures have a novel origin as condensation of 
topological kink dipoles (a condensation of $\langle iB_{i1}B_{i2}\rangle$ on 
each zz-bond).  On the other hand, field-induced {\it metamagnetic} transitions, now associated 
with `excitonic solid' states, are also generically expected from a 
Falicov-Kimball model with hybridization.
 
We analyze $H$ within dynamical mean-field 
theory (DMFT). Our choice is motivated by features special for 
$H_{K}$ : (i) when $h_{z}=0$, the KM possesses ultra-short-range spin 
correlations~\cite{baskaran}
rigorously uncorrelated beyond nearest 
neighbors, (ii) a DMFT-related approach~\cite{moessner} thus gives the first instance of an 
exact dynamical spin fluctuation spectrum in a $D=2$ model, and (iii) the fact 
that spin correlations on {\it all} length scales along $xx,yy$ bonds 
have been {\it exactly} subsumed into one-fermion hopping and $p$-wave BCS pairing 
terms obviates the usual difficulty of ignoring long-range spatial correlations, 
wherein any mean-field theory, including DMFT, is usually critiqued in $D=2$. Moreover, 
$H_{z}$ also has a local excitonic form and 
the $zz$-couplings appear as local interaction. Thus, these novelties, again specific 
for the KM, make the DMFT almost exact and allows for exhaustive analysis.

\vspace{0.20cm}

\begin{figure}
\centering
\subfigure[]{\label{f:C11}\epsfig{file=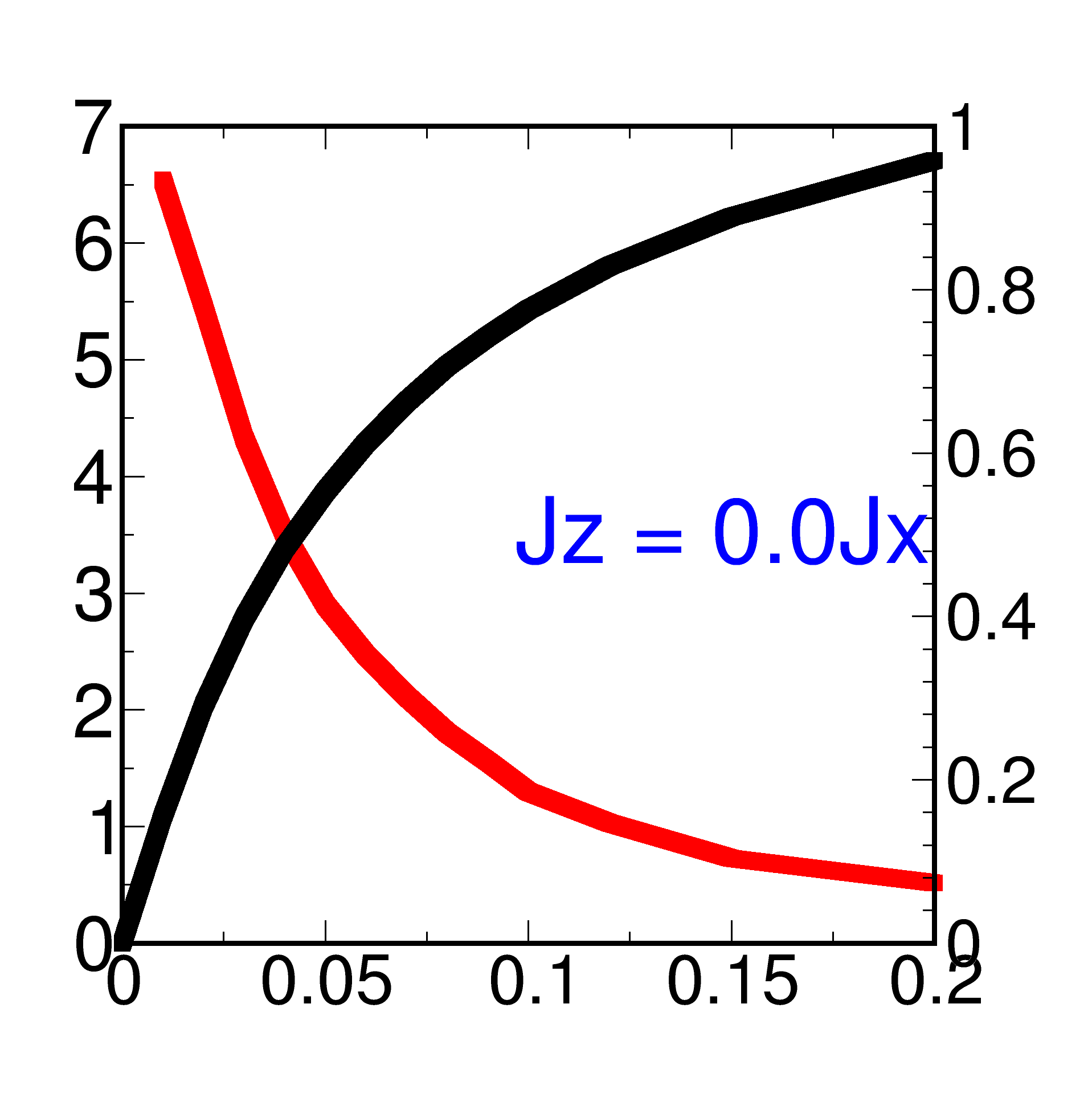,trim=0in 0in 0in 0.0in,
clip=true,width=0.32\linewidth}}\hspace{-0.0\linewidth}
\subfigure[]{\label{f:C21}\epsfig{file=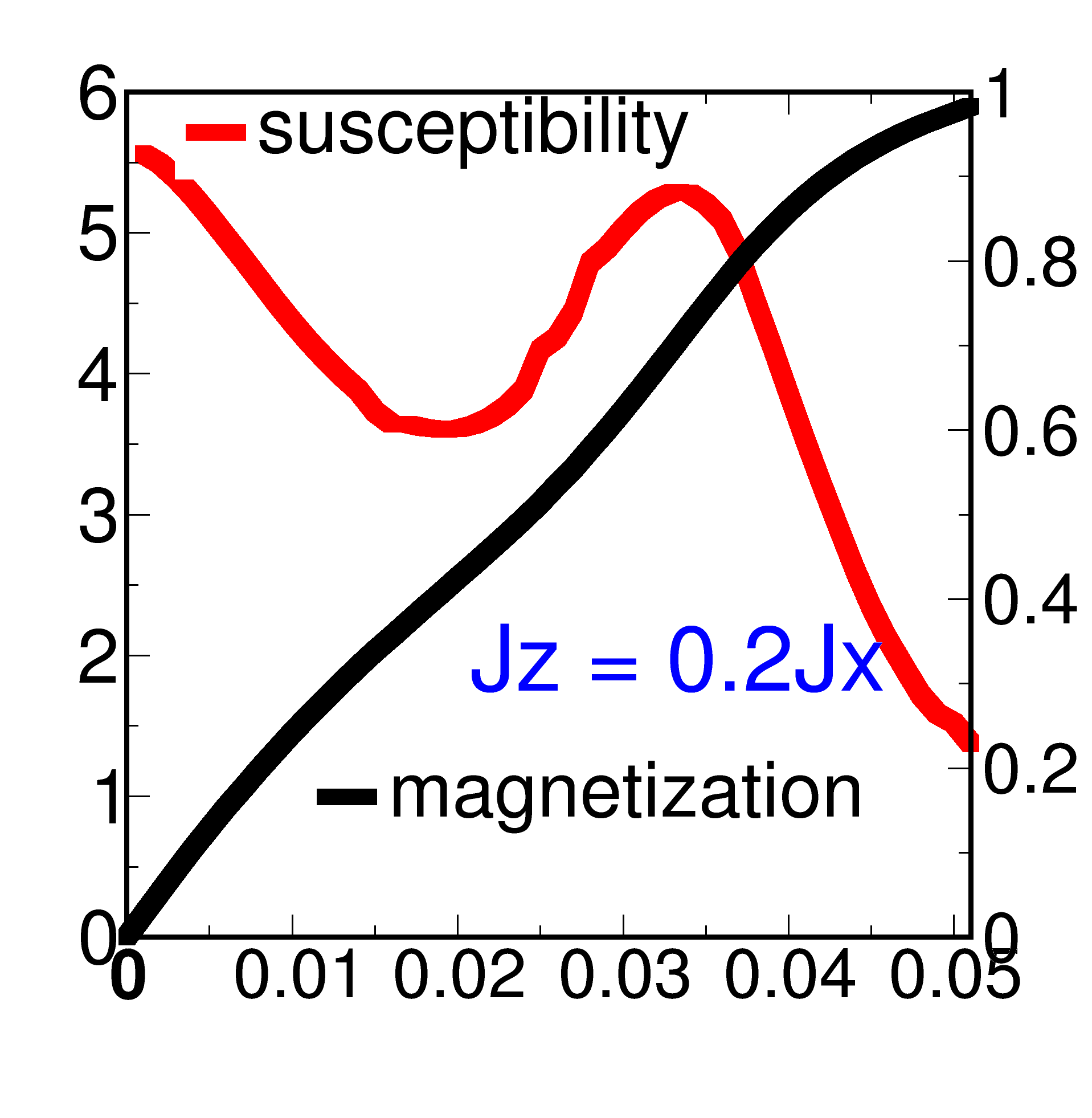,trim=0in 0in 0in 0.0in,
clip=true,width=0.32\linewidth}}\hspace{-0.0\linewidth}
\subfigure[]{\label{f:C21}\epsfig{file=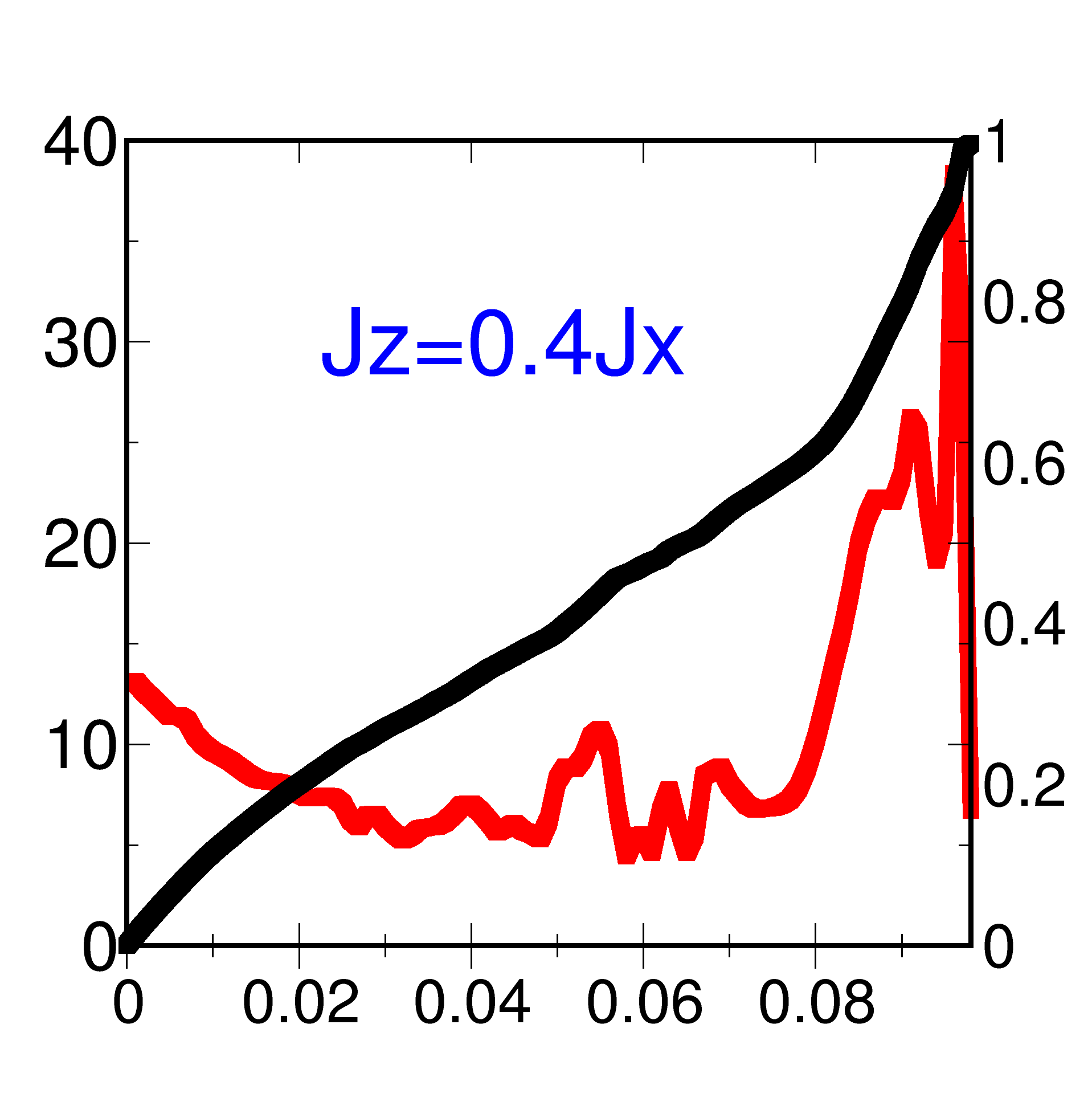,trim=0in 0in 0in 0.0in,
clip=true,width=0.32\linewidth}}\hspace{-0.0\linewidth}\\
\subfigure[]{\label{f:C11}\epsfig{file=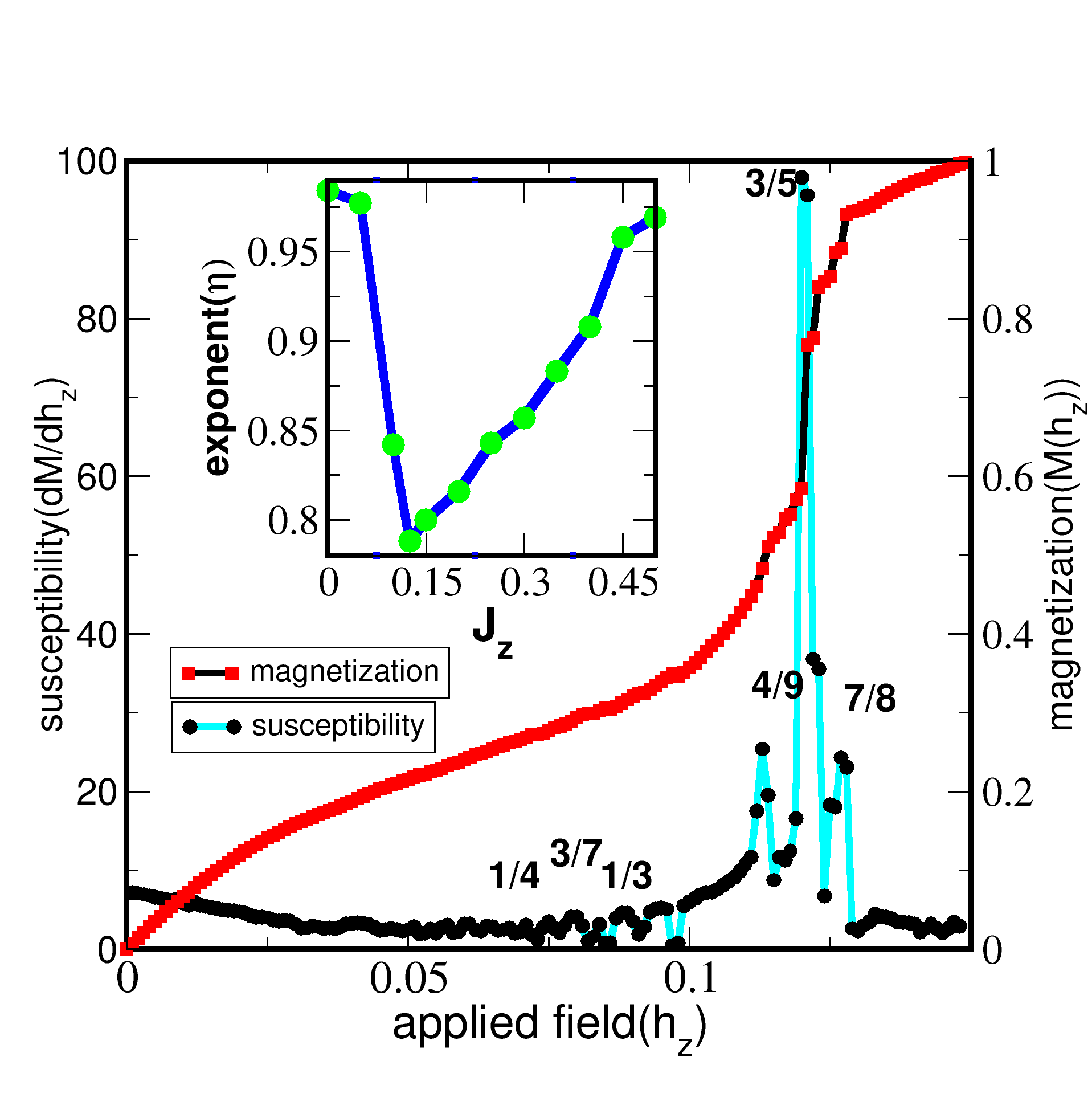,trim=0in 0in 0in 0.0in,
clip=true,width=0.82\linewidth}}\hspace{-0.0\linewidth}
\caption{Upper panel: Magnetization and static susceptibility as a function of externally applied field for
(a) $J_{z}=0$, (b) $J_{z}$=0.2$J_{x}$, and (c) $J_{z}$=0.4$J_{x}$. 
Lower panel: Plateaus and jumps in the anisotropic limit of the KM, with $J_{z}=0.6J_{x}$ (d). 
In the inset the $\eta$ is shown as a function of $J_{z}$ ($J_{x}=0.5$). The axes labels for figures (a), (b) and (c) are same as 
(d).} 
\label{fig2}
\end{figure}

 The impurity model of DMFT is solved by multi-orbital 
iterated perturbation theory as an impurity solver: though not ``exact'', its proven quantitative
accuracy vis-a-vis exact QMC results for the spinful Anderson 
lattice model~\cite{jarrell} ensures its efficacy in this case. We focus on
the $c,\alpha$-fermion spectral functions and field-induced magnetization, 
now simply equal to a local excitonic average $m(h_{z})=\langle (c_{i}^{\dag}\alpha_{i}+h.c)\rangle$. 
To facilitate discussion, we show the evolution of $m(h_{z})$ 
vs. $h_{z}$ in Fig.~\ref{fig2}. When $J_{z}<< J_{x}=J_{y}$, $m(h_{z})$ rises
monotonically with $h_{z}$, as expected of a field-polarized paramagnetic phase. 
However, its underlying topological origin is underscored by 
observing that $m(h_{z})\simeq h_{z}^{\eta}$ with $1.0 >\eta >0.85$ for 
$0\leq J_{z} \leq 0.25(=J_{x}/4)$: for larger $J_{z}$, $\eta(J_{z})$ reverses 
its slope (Fig.~\ref{fig2} lower panel (inset)), increasing monotonically toward $1.0$. The special 
character of $J_{z}/J_{x}=1/4$ is related to its being precisely twice
the (mean-field) excitation energy of the three flat bands that arise in 
a mean-field theory (related to the original resonating valence bond (RVB) description in the cuprate 
context~\cite{pwa}) distinct to ours~\cite{fiona}. It marks 
the point where the $\alpha$-fermion level and concomitant excitonic 
tendency begins to play a role for $J_{z}/J_{x}\geq 1/4$.  That $m(h_{z})$ starts developing 
plateaus (or incompressible JW-excitonic solid phases) as a function of 
$h_{z}$ {\it only} for $J_{z}/J_{x}>1/4$ also supports this view.\newline 

\vspace{-0.5em}
\begin{figure}
\centering
\includegraphics[angle=0,width=0.9\columnwidth]{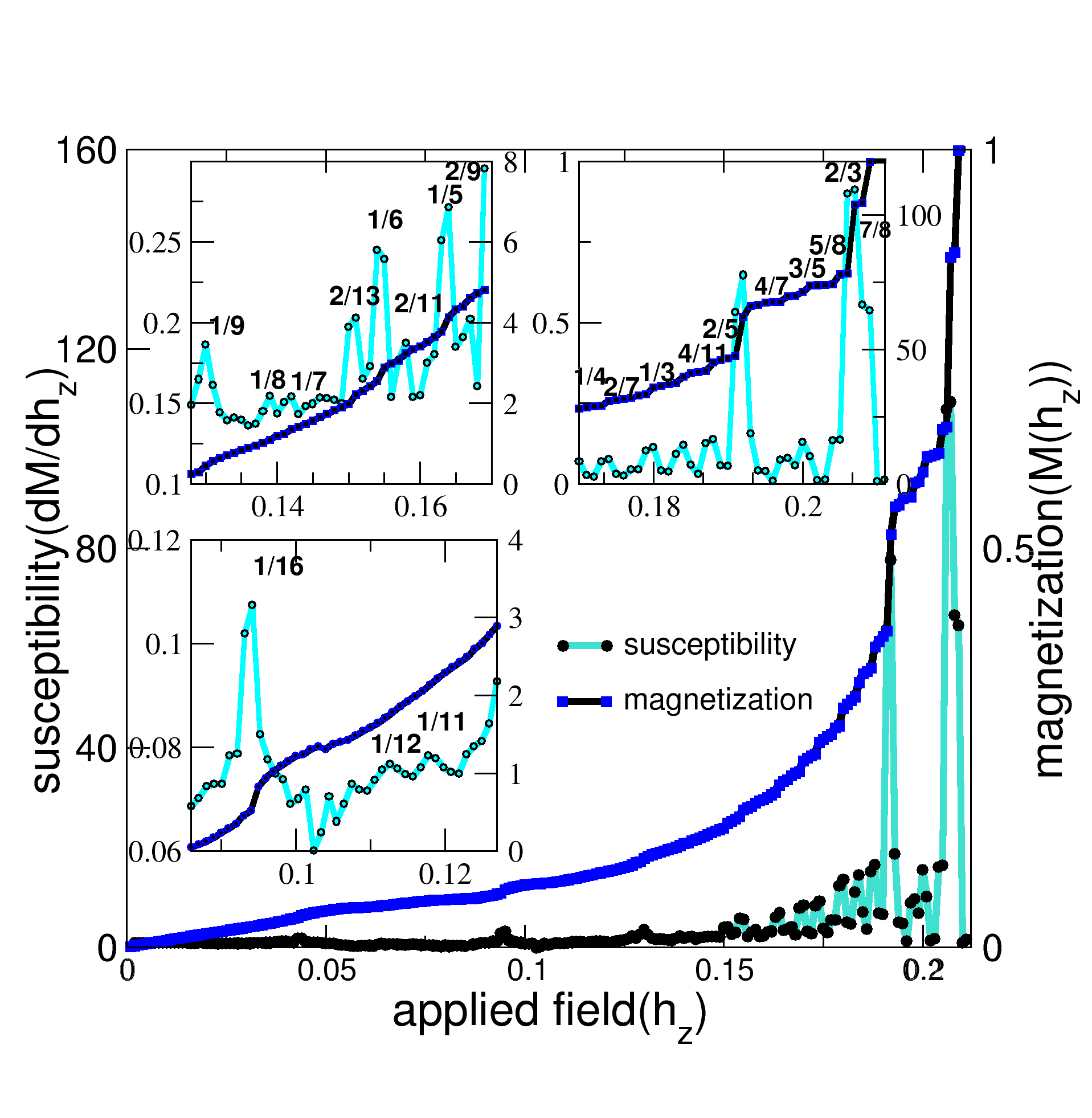}
\caption{Plateaus and jumps in the `magnetization' in the isotropic limit of
the KM. The corresponding susceptibilities are also shown.}
\label{fig3}
\end{figure}
  Remarkably, for $J_{z}/J_{x}>1/4$, $m(h_{z})$ as a function of $h_{z}$ 
shows an extensive sequence of magnetization plateaus, now associated with 
incompressible excitonic solid phases, separated by excitonic ``fluid'' phases.  These are intimately linked to the kink-dipole ``condensates'' (crystals of the
 ``flux'', $\langle iB_{i1}B_{i2}\rangle$: see the $n_{\alpha}(h_{z})$ vs $h_{z}$ in Suppl. Info).  
These are thus first-order quantum phase transitions between sequences of 
``topological crystal'' states, each sandwiched between topological 
Bose-Einstein condensates of excitonic kink dipoles.  These are nothing but zeeman-field induced metamagnetic transitions with accompanying hysteresis at each magnetization jump (see Fig.(3b) of supplementary informations),
and go hand-in-hand with jumps in $\langle n_{\alpha}(h_{z}\rangle$ at the same $h_{z}$ values.  These topological crystal phases occur (Fig.~\ref{fig3}) for a range of odd-denominator 
valued magnetizations, as well as at certain even denominators 
(relative to saturation, at $m(h_{z})/m_{sat}=1/16,1/12,1/11,1/9,1/8,1/7,2/13,1/6, 2/11, 1/5, 2/9, $\, $ 1/4,
 2/7, 1/3, 3/8, 3/7, 2/5, 4/7, 5/8, 3/4,$...). Odd denominator plateaus in the off-diagonal conductivity are well-known in the fractional quantum Hall effect (FQHE), which is 
also the {\it only} 
other example of a real system rigorously exhibiting topological order.  
In spin systems, such exotica have also been found and widely discussed
in frustrated Shastry-Sutherland models~\cite{mila-book, starykh}, where they emerge due to 
competition between frustration and field-induced magnetism (even-denominator plateaus are also visible there).  So our plateaus are not related to FQHE-like physics, even though (in contrast to earlier work) they arise in a perturbed (field-induced) partially TO phase.   Nevertheless, in Fig.~\ref{fig3}, we exhibit $m(h_{z})$ and $\chi(h_{z})=(d/dh_{z})m(h_{z})$ together to illustrate ``analogies'' with FQHE.  We interpret the oscillations in $\chi(h_{z})$ as ``de Haas van-Alphen (dHvA)'' quantum oscillations of topological JW fermions in a partially magnetized spin liquid phase.  Such exotica may show up~\cite{watanabe} in magnetic torque studies.  This reflects ``hidden'' coherence in a spin liquid: interestingly, hidden off-diagonal long-range order (ODLRO) was conjectured by early on by Anderson precisely in this context\cite{pwa-rvb}.  Here, $h_{z}\neq 0$ partially polarizes the spinons reflecting in magnetization plateaus, and these ordered phases `` ride on top'' of a partially topological ordered state.  This component has in-built coherence, thanks to the $p$-wave BCS term in $H_{K}$, manifesting as coherent ``quantum oscillations'' we unearth.  Thus, in stark contrast to conventional paramagnets, the TO phase of the Kitaev model undergoes an intricate sequence of partial ordering ``solid'' instabilities, co-existing with the remnant of the TO fractionalized spin liquid in an external Zeeman field before reaching full saturation.             

\vspace{0.1em}
\begin{figure}[ht!]
\centering
\subfigure[]{\label{f:C11}\epsfig{file=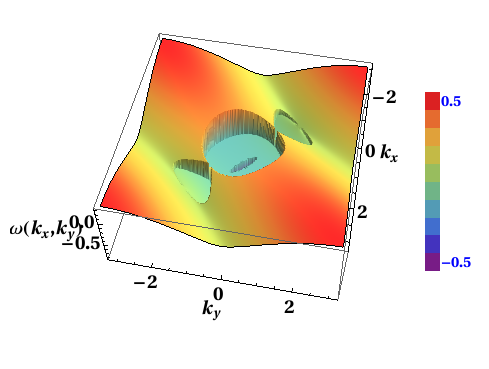,trim=0in 0in 0in 0.0in,
clip=true,width=0.75\linewidth}}\hspace{-0.0\linewidth}
\subfigure[]{\label{f:C21}\epsfig{file=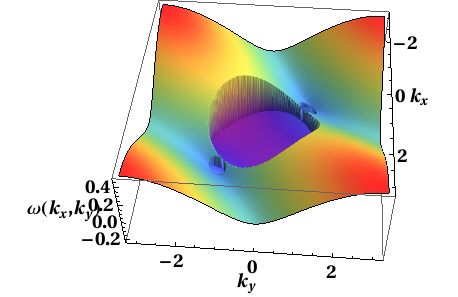,trim=0in 0in 0in 0.0in,
clip=true,width=0.75\linewidth}}\hspace{-0.0\linewidth}
\subfigure[]{\label{f:C21}\epsfig{file=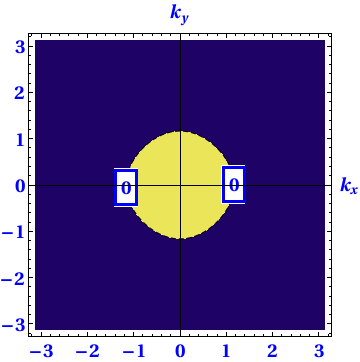,trim=0in 0in 0in 0.0in,
clip=true,width=0.48\linewidth}}\hspace{-0.0\linewidth}
\subfigure[]{\label{f:C21}\epsfig{file=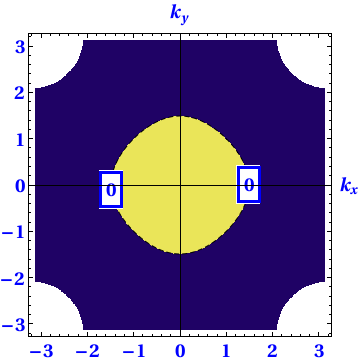,trim=0in 0in 0in 0.0in,
clip=true,width=0.48\linewidth}}\hspace{-0.0\linewidth}
\caption{The renormalized dispersions ((a) and (b)), corresponding Luttinger surfaces ((c) and (d)) for 
dispersive JW Fermions at plateaus 1/3 and 2/3 respectively. 
The boundary of yellow and blue regions corresponds to the contour of $\omega(k_{x},k_{y})=0$ 
and has been marked similarly inside the figures in the lower panel. 
$k_{x}$ and $k_{y}$ are in units of $\pi$/a, where a in lattice constant.}
\label{fig4}
\end{figure}

  The DMFT spectral functions reveal deeper insights (See supplementary informations).  Clear ``orbital'' selectivity is seen in 
$c$ and $\alpha$-fermion spectral functions, and this is a direct fall-out of the in-built differentiation between 
these states in the JW-fermionized $H_{K}$ itself.  While hybridization and local Hubbard correlations conspire to 
produce a JW ``Kondo'' insulator for $\alpha$-fermions, they produce large-scale spectral changes in the $c$-fermion 
sector.  Focussing on the variations in $\rho_{c}(\omega),\rho_{\alpha}(\omega)$ across representative samples where
magnetization steps appear, it is clear that large-scale reshuffling of dynamical spectral weight across energy scales of order the band-width (equal to $2J_{x}$ here) accompany emergence of each plateau

  Thus, the plateau structure has its 
origin in the ubiquitous competition between Mott localization (Hubbard term) and itinerance (hybridization and hopping)
in $H$, and the field simply tunes the balance between these tendencies, now by favoring the tendency to exciton formation.  This is reminiscent of the state of affairs characterizing the FL$^{*}$ view proposed in the context of $f$-electron quantum criticality~\cite{subir}.

\vspace{0.0cm}

  More unanticipated features show up in DMFT ``band structures''.  In Fig.~\ref{fig4}, we exhibit the $c$-fermion dispersions and ``Fermi'' surfaces obtained as usual from the DMFT propagators.  Remarkably, we find that the ``Fermi surface''  corresponds to {\it zeros}, rather than poles, of $G_{cc}(k,\omega)$: this is evident from the structure of $G_{cc}(k,\omega)$ itself, which features an anomalous ``BCS''-like self-energy structure given by $\Sigma_{an}(k,\omega)=\frac{\Delta_{k}^{2}}{\omega +\epsilon_{-k}-h^{2}/\omega}$, so that the {\it poles} of $\Sigma_{an}$ appear as zeros of $G_{cc}(k,\omega)$.  Thus, this is a generalized Luttinger surface, rather than a conventional Fermi surface.  This is analogous to what happens in correlation-driven selective-Mott transitions in $f$-electron systems in the context of the FL$^{*}$ phase, or in the YRZ ansatz~\cite{rice}.  In our case, clear changes in topology of the Luttinger surface occur at each plateau, and are thus intimately linked to correlation-driven Lifshitz-like transitions accompanying each field-induced plateau.  Moreover, given that the $c$-fermion spectral function is that of $p$-wave {\it nodal} Bogoliubov quasiparticles, this this state is a $p$-wave nodal {\it spin} semi-metal.  These novelties are signatures of the underlying partial TO surviving in the partially magnetized phases.  A Zeeman field results in the transfer of dynamical spectral weight from the fractionalized spin liquid component to the excitonic solid, but it is indeed remarkable, thanks to the undisturbed emergent $d=1$-GLS, that it partially preserves the original TO.   

  Finally, the large-$J_{z}$ limit of $H_{K}$ is known to be nothing but the Kitaev Toric Code (TC) model~\cite{kitaev}.
Here, this implies a {\it Mott}-like insulating phase of the JW fermions.  
The no double-occupancy constraint of a $c,\alpha$ fermion on each site $i$ is naturally implemented by a Gutzwiller projector, $P_{G}=\Pi_{i}(1-n_{ic}n_{i\alpha})$ acting on the $p$-wave BCS ground state.  Explicitly, 
$|\Psi\rangle_{TC} =P_{G}\Pi_{k}(u_{k}+v_{k}c_{k}^{\dag}c_{-k}^{\dag})|vac\rangle$ where $|vac\rangle$ is the $c$-fermion vacuum.  But this is precisely the Gutzwiller-projected ($p$-wave) BCS state, long known as one of the optimal
states describing a ($d$-wave) RVB spin liquid state in the context of the famed RVB description 
of the cuprates~\cite{pwa}.  Thus, the ground state of the TC model is just the $p$-wave RVB state!  This suggests the enticing possibility of investigating TC physics and its non-Abelian excitations in future in terms hitherto developed for quantum spin liquids in quantum Heisenberg models~\cite{sorella}.

\vspace{0.2cm}

  We now detail a specific quasi-realistic scenario where our findings could aid finding of novel orders in a real-world set-up:

{\bf 1. Kitaev Josephson Junction Arrays}    
As mentioned before (see also supplementary informations), a novel implementation of the perturbed KM using JJA has been proposed~\cite{jja}.  Our results can now be used to unearth novelties in such engineered systems.

\vspace{0.20cm}

   We now see that varying the flux through a qubit leads to a generation of a whole host of ordered ``JW-excitonic cooper-pair'' crystals, with a succession of periods set by the competition between $h_{z}$ (flux) and the material parameters of each JJ (these enter the Kitaev couplings~\cite{jja} in $J_{x},J_{y}=J,J_{z}$).  These are now interpreted as kink-dipole crystals, and arise as an explicit manifestation of the fact that strong geometric frustration-induced topological liquid state of the KM undergoes an intricate sequence of ``partially ordered'' solid phases before achieving a full solid at saturation.  In view of the co-existing and {\it remnant} topological critical 
spin-liquid arising from emergent ``$d=1$'' GLS in this state, one may also refer to each of these states as {\it critical} topological
supersolids: these turn out to be unstable to nematic order $\forall J_{x}\neq J_{y}$ (see supplementary informations).  A direct manifestation of this insight is that, in analogy with excitonic Josephson effects in electron-hole
bilayers~\cite{joglekar}, one expects the Josephson critical current, $J_{c}$, to depend on the excitonic correlations.
Explicitly, $J_{c} \simeq h_{z}^{2}\langle c_{i}^{\dag}\alpha_{i};c_{j}^{\dag}\alpha_{j}\rangle \simeq h_{z}^{2}\langle c_{i}^{\dag}\alpha_{i}\rangle\langle c_{j}^{\dag}\alpha_{j}\rangle$.  As a result of the $p$-wave ``BCS'' term in the fermionized $H_{K}$, there will also be a direct ``fermionic'' current, now caused by the $p$-wave JW-fermions.  Thus, remarkably, we expect the critical ``Josephson'' current to exhibit fractional
oscillations as the flux is slowly ramped up.  Fractional Shapiro steps are also directly predicted as a consequence of the underlying superstructure of kink-dipoles.  Inducing transitions between successive steps with a pulsed electromagnetic field can result in emission of coherent electromagnetic radiation whose energy $\hbar\Omega$ can be tuned in this setup.  With realistic JJ parameters, $\Omega$ can be tuned to lie in the GHz or sub-GHz region: thus, 
this setup can achieve flux-tuned control of EM radiation.  Excitations of such ordered states as those found here will be sliding collective modes of the excitonic solid akin to phason modes in the CDW context, and these are precisely the ``Josephson plasmons'', suggesting novel applications to ``topological plasmonics''.    
 Ours is thus a first proposal to exploit an almost exactly solvable model of fractionalized JW-fermions in $D=2$ to detect topological analogues of famed quantum interference effects.  Other subtle manifestations of ODLRO, such as topological analogues of Little-Parks experiment, etc, could also be imagined in the JJA context.  Further, to the extent that it is difficult to generate single JW fermions by splitting up JW excitons, the states we find have a much more efficient topological protection (the topological liquid phases are inherently symmetry-protected topological (SPT) phases), these findings can be exploited for possible topological quantum computation (TQC) engineering, along lines proposed by You {\it et al.}~\cite{jja}.

{\bf Conclusion}

  To conclude, we have presented strong evidence linking a large sequence of ``excitonic ordered solid'' phases of the Zeeman-field driven KM to novel instabilities of a rigorously known TO phase.  These could be touted as topological analogues of the famous "Barkhausen steps" encountered in the magnetization process of a conventional field-driven magnet.  Here, however, they correspond to partially magnetically ordered complex unit-cell patterns {\it emerging} as partial ordered states of a fractionalized spin liquid, betraying their highly unconventional nature.  Suitably engineered JJAs as proposed~\cite{jja} could unearth the structures we propose by flux tuning, with novel applications to novel quantum interference phenomena, plasmonics and TQC.  Finally, if examples of such perturbed
KMs to real transition-metal oxides could be found, manifestations of the underlying TO emerging as lower-$d(=1)$ gauge-like symmetries, together with their intimate link to unconventional nematic order,
could reveal themselves under appropriate external perturbations.  This remains an enticing perspective for future work.   
  
{\bf Acknowledgement}

MSL wishes to thank R. Moessner for a helpful discussion and MPIPKS, Dresden support when this work was conceived.  He also thanks G. Baskaran for helpful discussions.
AT and SA would like to acknowledge IMSc, where this work started, for  
hospitality. SA would like to thank UGC (India) for a research fellowship. 


\clearpage

\pagebreak
\widetext
\twocolumngrid
\begin{center}
\textbf{\large Supplementary Informations}
\end{center}
\setcounter{equation}{0}
\setcounter{figure}{0}
\setcounter{table}{0}
\setcounter{page}{1}
\makeatletter
\renewcommand{\theequation}{S\arabic{equation}}
\renewcommand{\thefigure}{S\arabic{figure}}
\renewcommand{\citenumfont}[1]{S#1}

{\bf $D=2$ Duality Transformations}

 Kitaev's famed solution relies on mapping spin-$1/2$s to bilinears of
four $(b^{x},b^{y},b^{z},c)$ Majorana fermions
along the three ($xx,yy,zz$) bonds (Fig.~\ref{fig4}). We start with an alternative but equivalent
approach due to Chen and Hu~\cite{Hu} and Nussinov~\cite{nussinov} that exploits
a duality mapping of the spins via a $D=2$ Jordan-Wigner (JW) fermionization of
$H$. Remarkably, since this affords
an {\it exact} mapping between topological and classical orders in $D=2$ spin
systems, studying TO(s), their excitations and instabilities is transmuted
into the better-known language of well-known models hosting classical order(s)
characterized by symmetry-breaking. Thus,

\vspace{0.1em}

\be
S_{ij}^{+}=2[\Pi_{j'<j,i} S_{i'j'^{z}}] [\Pi_{i'<i} S_{i'j}^{z}] c_{ij}^{\dag}
\ee
and

\be
S_{ij}^{z}=(2c_{ij}^{\dag}c_{ij}-1)
\ee
and defining Majorana fermions on the two (``black'' and ``white'') sublattice sites of the honeycomb or brick-wall lattice as
$A_{w}=(c-c^{\dag})_{w}/i,B_{w}=(c+c^{\dag})_{w}$ and $A_{b}=(c+c^{\dag})_{b},
B_{b}=(c-c^{\dag})_{b}/i$, followed by the introduction of fermions $c=
(A_{w}+iA_{b})/2,c^{\dag}=(A_{w}-iA_{b})/2$.  Using these transformations, we get the fermionized form of $H=H_{K}+H_{z}$ used in the main text.  In $H$ as thus defined in the main text, $\alpha_{i}'=iB_{i1}B_{i2}$ is a
bilinear of Majorana fermions.  Since $[\alpha_{i}', H]=0$ for all $i$, it takes eigenvalues $\pm 1$.  Thus, we
define $\alpha_{i}'=(2n_{i,\alpha}-1)$, where $n_{i,\alpha}=0,1$ and $[n_{i,\alpha},H]=0$ for all $i$ as well,
preserving the local Z$_{2}$ symmetry .  Using this representation, the zeeman term maps onto the excitonic
term in JW-fermion language.

\vspace{0.20cm}

{\bf Non-singular Spin Fluctuation Spectrum}

  Here, we point out how the ``Dirac'' spectrum for $c$-fermions cuts-off the infra-red (X-ray edge) singularities in the Kitaev model with $h_{z}=0$.
In a normal Fermi liquid, the infra-red singular nature of the localized propagator (as well as of the ``excitonic'' propagator) can be traced back to the appearance of logarithmic divergences in the particle-hole polarization function.  This necessitates summing of infinite number of Feynman diagrams for any arbitrary coupling strength~\cite{mahan}, or ``summing the logs'', leading to a power-law singularity in dynamical response functions.  In the Kitaev model, however, two features mitigate against this: first, the immobile-fermion level lies at $-J_{z}/2$ rather than at $\omega=0$, and, second, the ``Dirac''-like DOS for the $c$-fermions
(thanks to the $p$-wave BCS term in $H_{K}$) cuts off the infra-red singularities.  Explicitly, writing
$G_{\alpha\alpha|^{(0)}(i\omega)}=\frac{1}{i\omega+J_{z}/2}$  and
$G_{cc}(k,i\omega)=\frac{i\omega}{(i\omega)^{2}-v^{2}k^{2}}$

near the `Dirac'' points, we find that the one-loop polarization function,

\be
\Pi^{(0)}(i\omega)=-\int\frac{d\nu}{2\pi}\int\frac{d^{2}k}{(2\pi)^{2}}\frac{i\nu}{(i\nu)^{2}-(vk)^{2}}\frac{1}{i\nu-i\omega +J_{z}/2}
\ee

  behaves as $\Pi^{(0)}(i\omega)\simeq (2\pi v^{2})^{-1}[(i\omega +J_{z}/2)$ln$((J_{z}/2-i\omega)/W)-W]$, where
$W\simeq J_{x}$ is a cut-off of order the spinon bandwidth.  The log-singularity is now cut-off by the factor
$(i\omega+J_{z}/2)$, which itself arises from the linear DOS of the non-interacting $c$-fermions.  Thus,
there is no infra-red singularity in this case, and, as expected from this line of argument, a RPA calculation
using the bare propagators indeed captures the essential features of the exact solution to a high accuracy~\cite{Moessner}.

  In the TO phase ($h_{z}=0$), the $c$-fermions thus experience a strongly
fluctuating ($Z_{2}$) potential, which `suddenly' switches between
$0,J_{z}$ during every hop of the $c$-fermions.  If the pairing
term were absent, this would constitute a complete reconstruction of the
entire Fermi sea due to the Anderson-Nozieres-de-Dominicis ``orthogonality
catastrophe'' (OC).  Along with the rigorous ultra-short spin correlations in $H_{K}$,
this is what makes the `local' approximation exact here.  But with $h_{z}=0$, $\epsilon_{q}$
has a Dirac-like form which cuts off the infra-red singular behavior and the
OC, since the DOS vanishes linearly at $\omega=0(=\epsilon_{F})$ as shown above; this
will {\it not} be true anymore when
$h_{z}$ is finite and Fermi pockets, corresponding to finite DOS at $\omega=0$,
are expected to appear in the $c$-fermion spectrum. However, this is still not
enough, as the local hybridization also produces finite recoil of the
$\alpha$-fermion: this must cut-off the X-ray-edge singularities
in spectral functions. But since any finite $h_{z}$ also leads to an
excitonic condensate at $T=0$, this provides an additional route to removal of
infra-red singularities by a (partial in our case) coherence-restoring instability to field-induced order.
This is precisely what we find (see main text).

\vspace{0.20cm}
{\bf Non-Trivial Topological Order with $h_{z} \neq 0$}

  Here, we show how the emergent $D=1$ GLS mentioned in the text are linked to a novel ``hidden'' topological order even when $h_{z} \neq 0$.

  Noticing that a Zeeman field along $zz$-bonds implies $[ib_{i}^{a}b_{j}^{a},H_{K}+H_{z}]=0$ for all $(ij)||xx,yy$ implies persistence of {\it partial} local $Z_{2}$ gauge symmetry and emergent low ($d=1$) gauge-like symmetries, we show that these hide a novel kind of topological quantum critical point (TQCP), separating two
topological phases.  Since topological order can be described by conventional (Landau-like) order when expressed in terms of dual (Jordan-Wigner) variables, we take this approach.  Remarkably, it will turn out that the special point $J_{x}=J_{y}(=J)$ separates two {\it Ising}-like nematic phases at $T=0$.

  In the single chain case, $H_{K}$ is trivially diagonalized to give two quasiparticle bands with dispersions $E_{k}^{\pm}=\pm\sqrt{J_{x}^{2}+J_{y}^{2}+2J_{x}J_{y}cos(k)}$ (with one band fully occupied and the other empty).  The second derivative of the ground state energy diverges at $J_{x}=J_{y}$, where the excitation gap vanishes.  This quantum transition at $T=0$ does not involve any symmetry change, rather, it is a change in {\it topological} order.  This is made explicit by the transformation~\cite{Feng,Susskind}
\vspace{0.1em}

\be
S_{i}^{x}=\tau_{i-1}^{x}\tau_{i}^{x},  S_{i}^{y}=\Pi_{l=i}^{2N} \tau_{l}^{y}
\ee
which recasts $H_{K}$ as the $D=1$ quantum Ising model (QIM) on the dual lattice.  When $J_{x}>J_{y}$the dual spin-correlation function shows long-range order:

\be
Lim_{i\rightarrow\infty}\langle\tau_{0}^{x}\tau_{2i}^{x}\rangle \simeq [1-(J_{y}/J_{x})^{2}]^{1/4}
\ee
In the original spin-representation, this is a {\it string} order with ``order'' parameter
$\Delta_{x}(i)=\langle\Pi_{l=1}^{2i}S_{l}^{x}\rangle$.

  For $J_{y}>J_{x}$, we exploit the well-known Kramers-Wannier duality of the $D=1$ QIM to conclude that
the order is that of strings of the $S^{y}$, i.e, the order parameter is $\Delta_{y}=\langle \Pi_{l=2}^{2i+1}S_{l}^{y}\rangle$.

  At $J_{x}=J_{y}$, both these order parameters collapse to zero simultaneously.  The topological excitations which destroy each of these orders are precisely proliferated string-like excitations associated with the {\it dual} order: these are consequences of the emergent $d=1$ GLS.  Remarkably, it is easily seen that the two non-zero order parameters are nothing but
the {\it only} ones invariant under the emergent GLS, i.e, they are the spin-{\it nematic} order parameters,
defined by $\langle n\rangle=\langle S_{i}^{x}S_{j}^{x}-S_{i}^{y}S_{j}^{y}\rangle$.  Thus, $J_{x}=J_{y}$ is a TQCP separating two topological ordered phases.  In dual language, it separates two Ising-nematic ordered
phases.

\vspace{0.50cm}
\vspace{0.1em}
\begin{figure}
\centering
\includegraphics[angle=0,width=0.7\columnwidth]{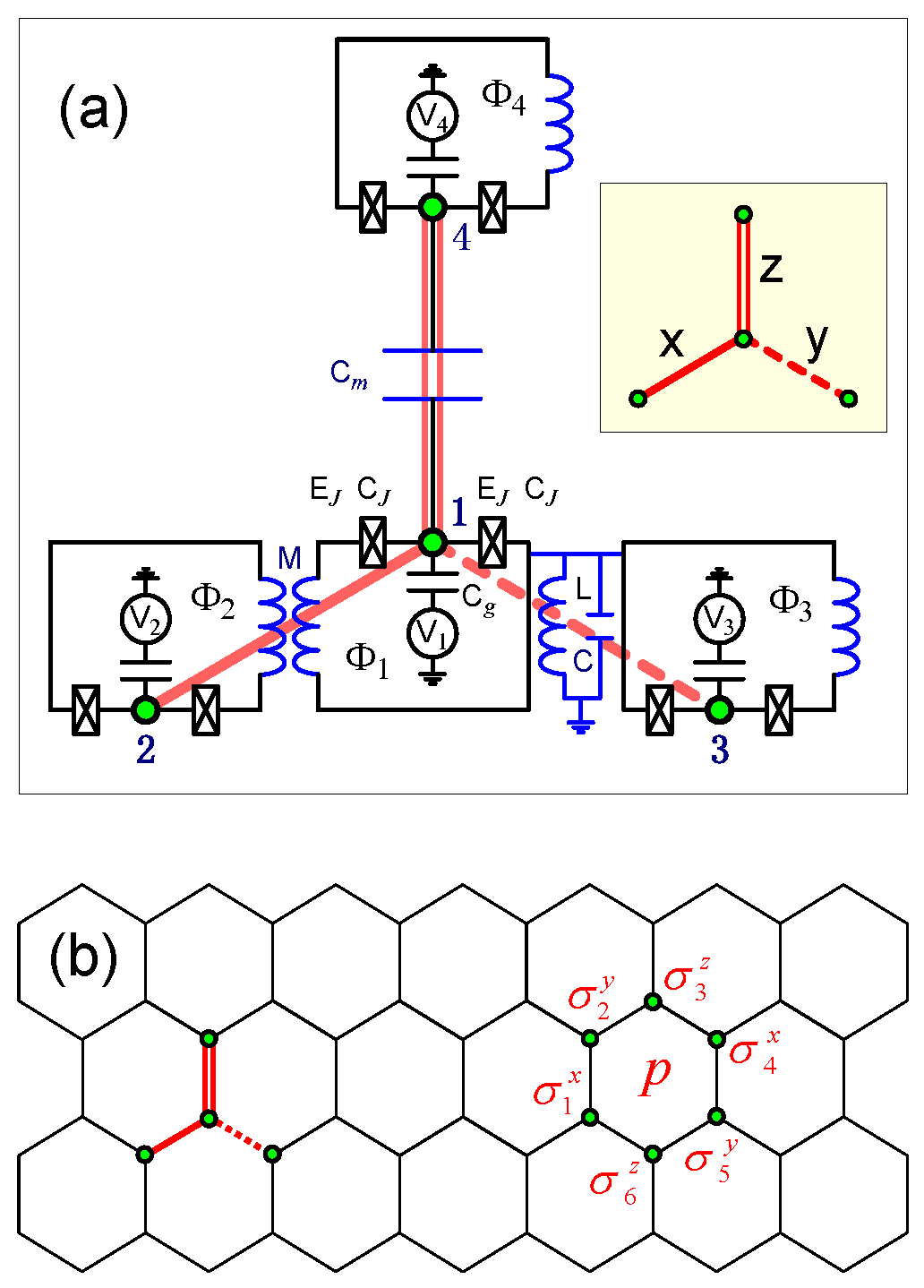}
\caption{(Color online) JJA on a honeycomb lattice(~\cite{JJA}).}
\label{fig4}
\end{figure}
{\bf Implementation of the Perturbed Kitaev Model in Engineered Josephson Junction Arrays}

  Here, we briefly recapitulate the proposal for simulating the perturbed Kitaev model we have studied by using a specific engineered Josephson junction array on the honeycomb lattice.

 Our results are directly applicable to the engineered JJA of You {\it et al.}~\cite{JJA} as long as we identify the Kitaev spins as pseudospin-$1/2$s associated with a pair charge qubit: defining $|\uparrow\rangle_{i},|\downarrow\rangle_{i}$ as two charge states with cooper pair number $n_{i}=1,0$, the $S=1/2$ representation permits a rigorous mapping, $n_{i}=(1-S_{i}^{z})/2,$ cos$\phi_{i}=S_{i}^{x}/2$ and sin$\phi_{i}=-S_{i}^{y}/2$ with $\phi_{i}$ being the pair phase at $i$, conjugate to the number as implied by $n_{i}=-id/d\phi_{i}$.  If each charge qubit is placed at an ``optimal'' point where $n_{g}=1/2$, and $J_{x}(=J_{y})<J_{z}$ is considered, one ends up with the anisotropic Kitaev model in an ``external field'' $h\sum_{i}S_{i}^{x}$ ({\it not} $hS_{i}^{z}$).  However, this is of no consequence, as we obtain our model simply by relabelling $S_{i}^{x}\rightarrow S_{i}^{z}, S_{i}^{z}\rightarrow -S_{i}^{x}$.  Further, we then have $J_{z}<J_{x}(=J)$, which is precisely the limit we are concerned with.
 Remarkably, since $h=E_{Ji}(\Phi_{i})=2E_{J}$cos$(\pi\Phi_{i}/\Phi_{0})$ with $\Phi_{0}=\pi\hbar/e$ the flux quantum in
the JJA, changing the Josephson coupling energy offers a subtle knob to tune the ``Zeeman'' field in the KM, and, in fact, in the JJA context, a real external magnetic field or adding non-magnetic impurities can change $E_{Ji}(\Phi)$~\cite{Tsvelik}.

\vspace{0.1em}
\begin{figure}[ht!]
\centering
\subfigure{\label{f:C11}\epsfig{file=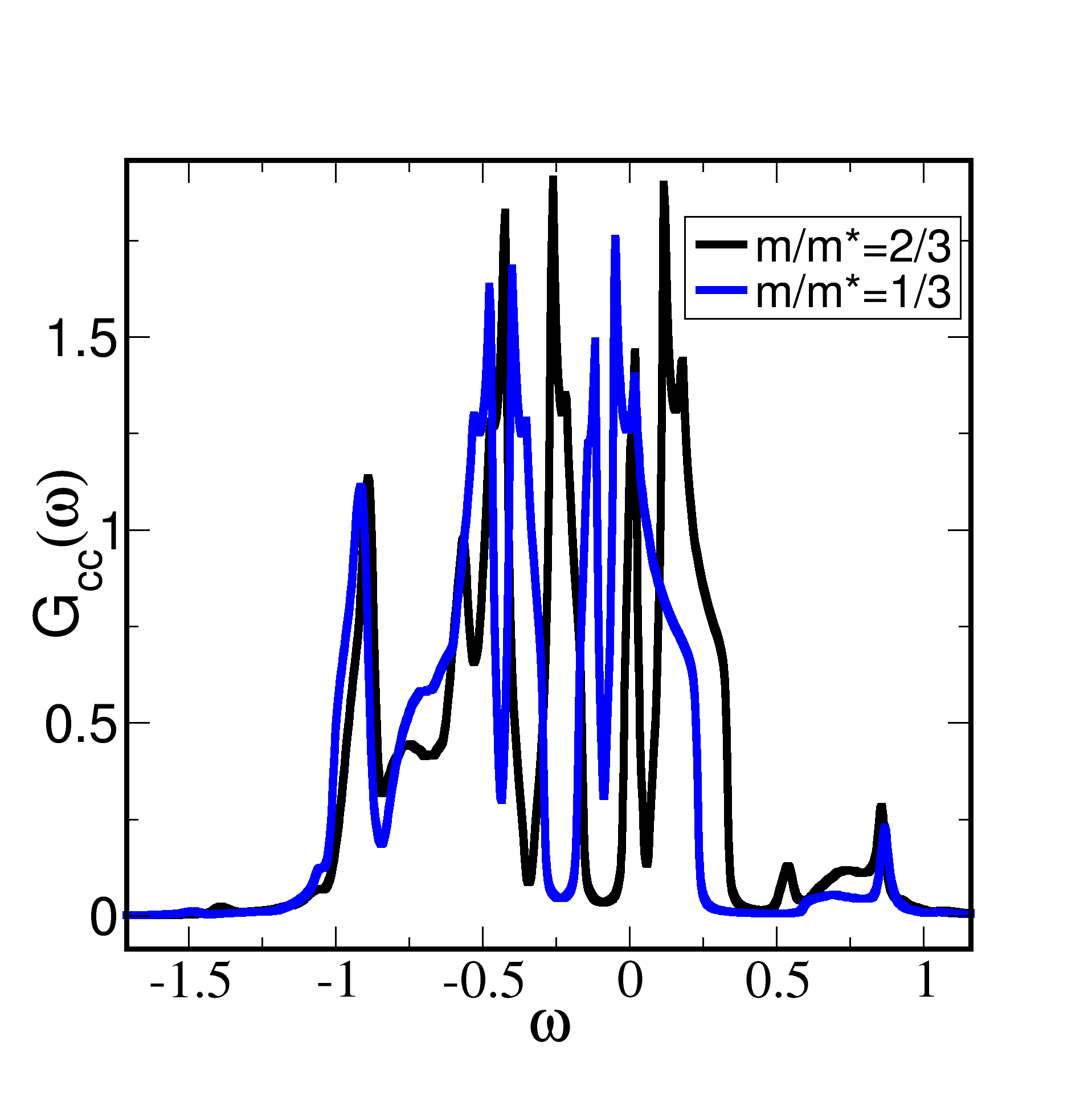,trim=0in 0in 0in 0.0in,
clip=true,width=0.49\linewidth}}\hspace{-0.0\linewidth}
\subfigure{\label{f:C21}\epsfig{file=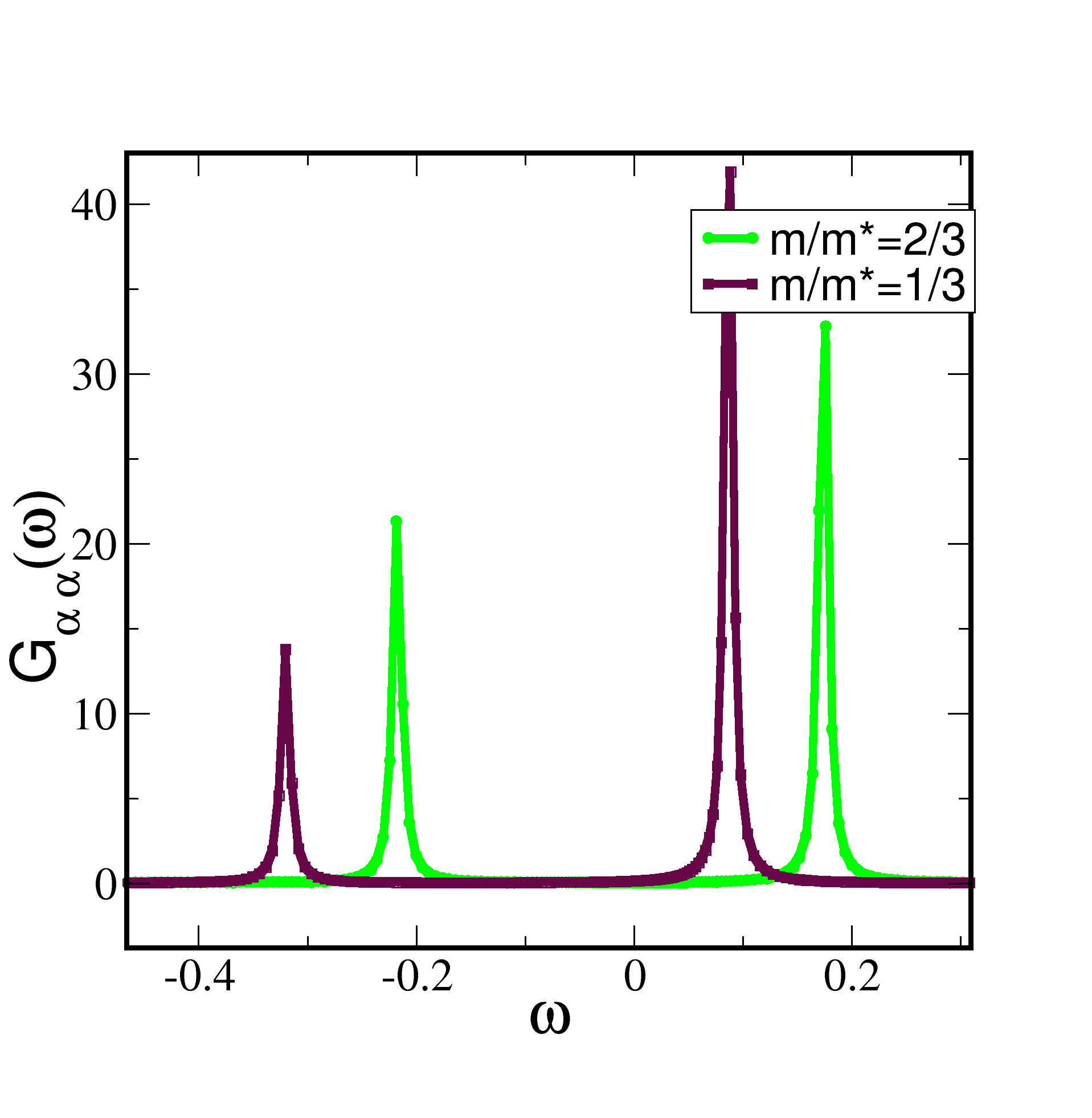,trim=0in 0in 0in 0.0in,
clip=true,width=0.49\linewidth}}\hspace{-0.0\linewidth}\\
\caption{(Color online) The DOS of the interacting $c$ and $\alpha$ (left and right respectively) fermions at plateaus $\frac{1}{3}$ and $\frac{2}{3}$ .}
\label{fig5}
\end{figure}

\vspace{0.20cm}

{\bf DMFT Spectral Functions}

  Here, we expound briefly upon our DMFT results for the spectral functions used to discuss spectral features and their consequences in the main text.

  We have used the two-orbital iterated perturbation theory (IPT) as an impurity solver in the context of solving $H$ using DMFT.  Though not ``exact'' in the sense of continuous-time quantum Monte Carlo (CT-QMC) or the numerical renormalization group (NRG) impurity solvers, its numerical efficacy in the context of DMFT for a related {\it spinful} Anderson lattice problem vis-a-vis CTQMC is
well-documented, allowing a good degree of
confidence for its use in our case.

\vspace{0.1em}
\begin{figure}
\centering
\subfigure[]{\label{f:C11}\epsfig{file=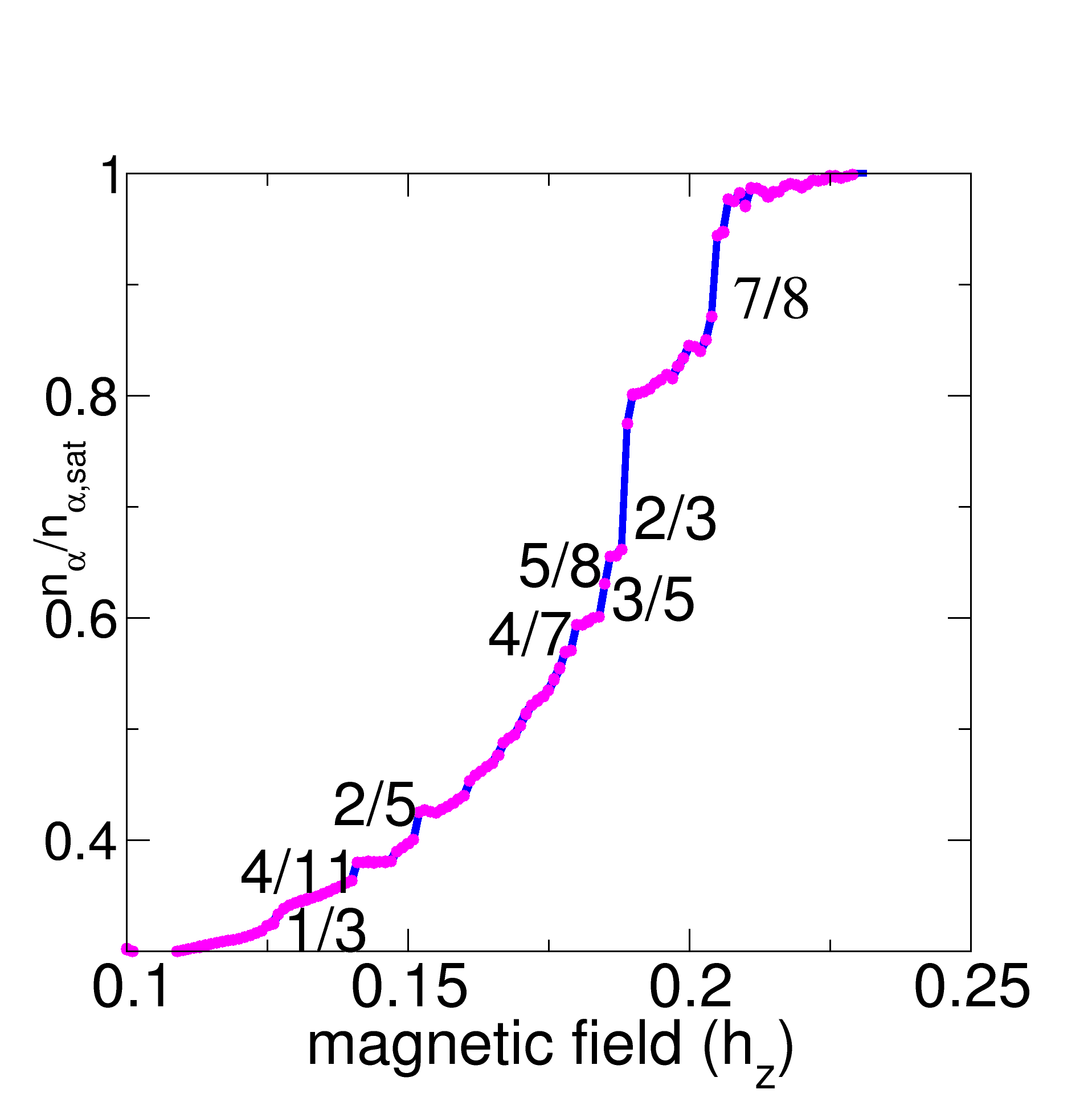,trim=0in 0in 0in 0.0in,
clip=true,width=0.49\linewidth}}\hspace{-0.0\linewidth}
\subfigure[]{\label{f:C21}\epsfig{file=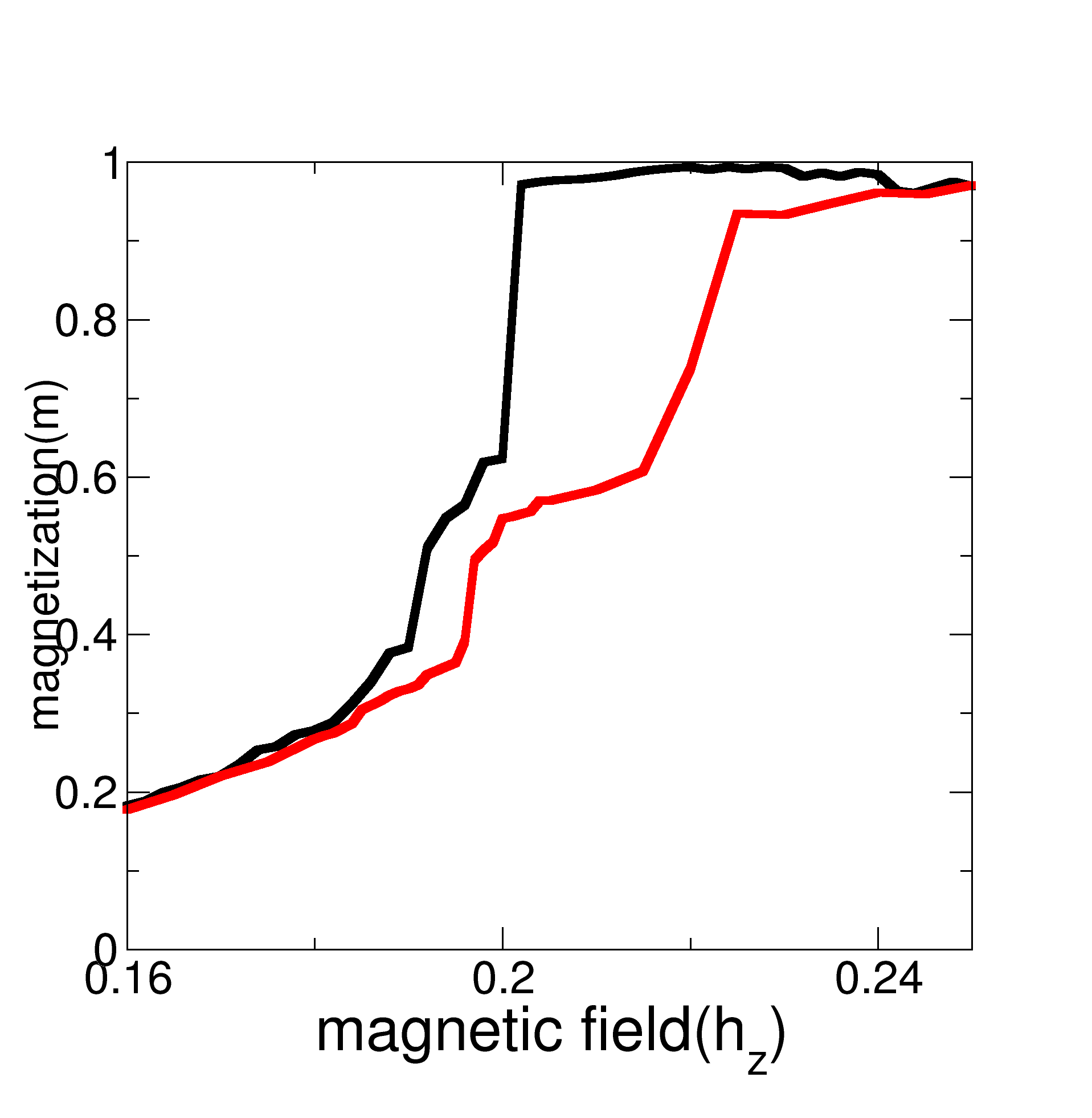,trim=0in 0in 0in 0.0in,
clip=true,width=0.49\linewidth}}\hspace{-0.0\linewidth}\\
\caption{(Color online)(a) $n_{\alpha}(h_{z})/n_{\alpha,sat}$ as a function of magnetic field ($h_{z}$) in the isotropic limit ($n_{\alpha,sat}$ is the saturation value of $n_{\alpha}$). The jumps and plateaus in the profile closely correspond to the same
numbers in $m(h_{z})/m_{sat}$ in the same limit. (b) Hysteresis in magnetization.}
\label{fig6}
\end{figure}

  In Fig.~\ref{fig5}, we show the $c$- and the $\alpha$-JW fermion spectral functions for the isotropic Kitaev case at values $m(h_{z})/m_{sat}=1/3,2/3$.  Several notable features stand out clearly.  First, the in-built
``orbital'' differentiation involved in the pure KM has direct consequences on their respective evolution as a function of $h_{z}$.  In particular, the $\alpha$-fermion DOS resembles that of a Kondo insulator, owing to the
fact that a combination of $h_{z}$ induced $c-\alpha$ hybridization and the bare energy level of the $\alpha$-fermion in the presence of a ``Hubbard'' ($J_{z}$) interaction term conspire to generate such a state for the
parameter values set by $H$.  However, the $c$-fermion DOS shows drastic spectral changes as a direct consequence of the dynamical interplay between the ``Dirac'' band structure of the $c$-fermions, $c-\alpha$ hybridization, and the Hubbard interaction.  Notwithstanding this, however, the $c$-JW fermions continue to be
elementary excitations of the perturbed spin liquid, since Im$\Sigma_{cc}(\omega)$ always vanishes at low energy in DMFT (not shown).  This aspect is a direct consequence of the important observation, namely, that a finite $h_{z}$ leads to {\it emergent} lower-dimensional $d(=1)<D=2$ gauge-like symmetries: thus, these spinon-like excitations are topological excitations (non-local strings in the original spin variables in $H_{K}$) of this partially TO state.  However, this fractionalized component co-exists with a finite ``excitonic'' order
parameter, which is just the field-induced magnetization (Fig.~\ref{fig6}(a)) in original spin language.

   This has important consequences:  clearly, our results show that,
across successive magnetization plateaus, large reshuffling of dynamical spectral weight over scales of order the full $c$-fermion band-width occur.  This clearly attests to the important role of dynamical correlations
in this interplay and generation of the plateau structures.  To the extent such dynamical correlations are crucial, we opine that less sophisticated approximations such as static Hartree-Fock may miss important aspects of the detailed structures found here.

\end{document}